\documentclass{PoS}
\usepackage{cite}

\newcommand{\bq}{\begin{eqnarray}}
\newcommand{\eq}{\end{eqnarray}}
\newcommand{\eps}{\varepsilon}

\title{A walk on sunset boulevard}

\ShortTitle{A walk on sunset boulevard}

\author{Luise Adams\\
        Johannes Gutenberg-Universit\"at Mainz\\
        E-mail: \email{ladams01@uni-mainz.de}}

\author{Christian Bogner\\
        Humboldt-Universit\"at zu Berlin\\
        E-mail: \email{bogner@math.hu-berlin.de}}

\author{\speaker{Stefan Weinzierl}\\
        Johannes Gutenberg-Universit\"at Mainz\\
        E-mail: \email{weinzierl@uni-mainz.de}}

\abstract{
A walk on sunset boulevard can teach us about transcendental functions associated to Feynman diagrams.
On this guided tour we will see multiple polylogarithms, differential equations and elliptic curves.
A highlight of the tour will be the generalisation of the polylogarithms to the elliptic setting
and the all-order solution for the sunset integral in the equal mass case.
}

\FullConference{12th International Symposium on Radiative Corrections (Radcor 2015) 
                and LoopFest XIV (Radiative Corrections for the LHC and Future Colliders)\\
                15-19 June  2015\\
                UCLA Department of Physics \& Astronomy Los Angeles, CA, USA}

\begin{document}

\section{Motivation}

Analytic calculations of Feynman integrals are important for precision particle physics.
Due to the presence of ultraviolet or infrared divergences these calculations 
usually employ dimensional regularisation.
The result is presented as a Laurent series in the dimensional regularisation parameter $\eps$.
This talk is centred around two questions:
\begin{center}
\begin{tabular}{ll}
Q1: & Which transcendental functions appear in the $\eps^j$-term?
\\
Q2: & What are the arguments of these function?
\end{tabular}
\end{center}
We are far away from giving a complete answer to these two questions.
However, there has been significant progress in the past years and in this talk we report on the state-of-the-art.

Let us start with the basics:
For one-loop integrals and for the expansion around four space-time dimensions the answer to question 1
for the $\eps^0$-term is simple: 
There are just two transcendental functions. These are the logarithm
and the dilogarithm
\bq
\label{def_logarithm}
 \mathrm{Li}_1\left(x\right) 
 \;\; = \;\;
 - \ln\left(1-x\right)
 \;\; = \;\;
 \sum\limits_{n=1}^\infty \frac{x^n}{n},
 & \hspace*{15mm} & 
 \mathrm{Li}_2\left(x\right) 
 \;\; = \;\;
 \sum\limits_{n=1}^\infty \frac{x^n}{n^2}.
\eq
There is a wide class of Feynman integrals which evaluate to generalisations of the two transcendental functions above,
called multiple polylogarithms. We review multiple polylogarithms in the next section.
The multiple polylogarithms are functions, which by now are well understood.

Beyond the class of multiple polylogarithms we encounter ``terra incognita''.
There are Feynman integrals, which cannot be expressed in term of multiple polylogarithms.
The simplest integral of this type is the two-loop sunset integral
(also known as sunrise integral in the eastern parts of the world).
For this reason, the two-loop sunset integral is a guide, which allows us to explore the ``terra incognita'' of 
functions beyond the class of multiple polylogarithms.

We may explore this field systematically step-by-step:
We first determine an (inhomogeneous) differential equation for the (yet) unknown two-loop sunset integral.
We then solve the differential equation: We first find the solutions of the corresponding homogeneous differential
equation and then construct the solution of the original inhomogeneous differential equation.
In all these steps guidance from algebraic geometry is very helpful.

\section{Multiple polylogarithms}

An obvious generalisation of the logarithm and the dilogarithm in eq.~(\ref{def_logarithm}) 
are the (classical) polylogarithms (with $m \in {\mathbb N}$):
\bq
\label{def_polylogarithm}
 \mathrm{Li}_m\left(x\right) & = & 
 \sum\limits_{n=1}^\infty \frac{x^n}{n^m}.
\eq
Explicit calculations teach us that we need in addition a generalisation to multiple arguments, which brings us 
to multiple polylogarithms.
The multiple polylogarithms are defined by \cite{Goncharov_no_note,Goncharov:2001,Borwein}
\bq
 \mathrm{Li}_{n_1,n_2,...,n_k}\left(x_1,x_2,...,x_k\right)
 & = &
 \sum\limits_{j_1=1}^\infty \sum\limits_{j_2=1}^{j_1-1} ... \sum\limits_{j_k=1}^{j_{k-1}-1}
 \frac{x_1^{j_1}}{j_1^{n_1}} \frac{x_2^{j_2}}{j_2^{n_2}} ... \frac{x_k^{j_k}}{j_k^{n_k}}.
\eq
The multiple polylogarithms have also a representation as iterated integrals.
Let us define functions $G$ for $z_k \neq 0$ by
\bq
\label{Gfuncdef}
G(z_1,...,z_k;y) 
 & = &
 \int\limits_0^y \frac{dt_1}{t_1-z_1}
 \int\limits_0^{t_1} \frac{dt_2}{t_2-z_2} ...
 \int\limits_0^{t_{k-1}} \frac{dt_k}{t_k-z_k}.
\eq
In this definition one variable is redundant due to the scaling relation $G(z_1,...,z_k;y) = G(x z_1, ..., x z_k; x y)$.
To relate the multiple polylogarithms to the functions $G$ it is convenient to introduce the following short-hand notation:
\bq
 G_{m_1,...,m_k}(z_1,...,z_k;y)
 & = &
 G(\underbrace{0,...,0}_{m_1-1},z_1,...,z_{k-1},\underbrace{0...,0}_{m_k-1},z_k;y).
\eq
Here, all $z_j$ for $j=1,...,k$ are assumed to be non-zero.
One then finds
\bq
 \mathrm{Li}_{m_1,...,m_k}(x_1,...,x_k)
 & = & 
 (-1)^k 
 G_{m_1,...,m_k}\left( \frac{1}{x_1}, \frac{1}{x_1 x_2}, ..., \frac{1}{x_1...x_k};1 \right).
\eq
Methods for the numerical evaluation of multiple polylogarithms are available \cite{Vollinga:2004sn}.
On the mathematical side, multiple polylogarithms are closely related to punctured Riemann surfaces of genus zero \cite{Goncharov:2001,Brown:2006,Bogner:2014mha}.

\section{Differential equations for Feynman integrals}

Let us consider a scalar Feynman integral. 
This integral may depend on Lorentz invariants $s_{jk}$ and internal masses squared $m_i^2$.
Suppose that it is not feasible to compute the integral directly.
A possible strategy is to split the task into two parts:
Let us pick one variable $t$ from the set $\{s_{jk},m_i^2\}$.
We first try to find an ordinary differential equation for the (unknown) Feynman integral $I_G(t)$:
\bq
  \sum\limits_{j=0}^r p_j(t) \frac{d^j}{dt^j} I_G(t) & = & \sum\limits_i q_i(t) I_{G_i}(t).
\eq
In general we will obtain an inhomogeneous differential equation, where the inhomogeneous 
term consists of simpler (known) integrals $I_{G_i}$.
The coefficients $p_j(t)$, $q_i(t)$ are polynomials in $t$.
The number $r$ denotes the order of the differential equation.
In a second step one tries to solve the differential equation.
It is always possible to perform the first step, so the non-trivial part consists in solving the differential equation.
Methods and algorithms for finding the differential equation 
can be found in \cite{Kotikov:1990kg,Kotikov:1991pm,Remiddi:1997ny,Gehrmann:1999as,Argeri:2007up,MullerStach:2012mp,Henn:2013pwa,Henn:2014qga}.

Let us look at a few special cases:
Suppose the differential operator factorises into linear factors:
\bq
 \sum\limits_{j=0}^r p_j(t) \frac{d^j}{dt^j}
 & = & 
 \left( a_r(t) \frac{d}{dt} + b_r(t) \right)
 ...
 \left( a_2(t) \frac{d}{dt} + b_2(t) \right)
 \left( a_1(t) \frac{d}{dt} + b_1(t) \right).
\eq
This corresponds to an iteration of $r$ first-order differential equations and can be solved step-by-step
with the methods for first-order differential equations. 
We denote the homogeneous solution of the $j$-th factor by
\bq
 \psi_j(t) & = & \exp\left(- \int\limits_0^t ds \frac{b_j(s)}{a_j(s)} \right).
\eq
The full solution of the differential equation is given by iterated integrals of the form
\bq
\label{iterated_solution}
 I_G(t)  & = &
 C_1 \psi_1(t) + \psi_1(t) \int\limits_0^t \frac{dt_1}{a_1(t_1) \psi_1(t_1)} 
                  \left( C_2 \psi_2(t_1) + \psi_2(t_1) \int\limits_0^{t_1} \frac{dt_2}{a_2(t_2) \psi_2(t_2)}
                              ...
 \right. 
 \nonumber \\
 & & \left. 
 ...
                                  \left( C_r \psi_r(t_{r-1}) + \psi_r(t_{r-1}) \int\limits_0^{t_{r-1}} \frac{dt_r}{a_r(t_r) \psi_r(t_r)} 
                                                         \sum\limits_i q_i(t_r) I_{G_i}(t_r)
                              \right) 
           \right).
\eq
The $r$ integration constants are denoted by $C_1$, ..., $C_r$.
From the integral representation of the multiple polylogarithms in eq.~(\ref{Gfuncdef}) we deduce that 
multiple polylogarithms are of this form.

We are interested in transcendental functions, which go beyond the class of multiple polylogarithms.
Suppose the differential operator 
\bq
 \sum\limits_{j=0}^r p_j(t) \frac{d^j}{dt^j}
\eq
does not factor into linear factors.
The next more complicated case consists of a  
differential operator which contains
one irreducible second-order differential operator
\bq
 a_j(t) \frac{d^2}{dt^2} + b_j(t) \frac{d}{dt} + c_j(t).
\eq
Let us first look at an example from mathematics.
The differential operator of the homogeneous second-order differential equation
\bq
 \left[ t \left(1-t^2\right) \frac{d^2}{dt^2} + \left(1-3t^2\right) \frac{d}{dt} - t \right] f(t) & = & 0
\eq
is irreducible.
The solutions of this differential equation are $K(t)$ and $K(\sqrt{1-t^2})$,
where $K(t)$ is the complete elliptic integral of the first kind:
\bq
 K(t)
 & = &
 \int\limits_0^1 \frac{dx}{\sqrt{\left(1-x^2\right)\left(1-t^2x^2\right)}}.
\eq
We will soon encounter irreducible second-order differential operators and elliptic integrals in a physics case.

\section{The two-loop sunset integral}

It is now time to introduce the two-loop sunset integral \cite{Broadhurst:1993mw,Bauberger:1994by,Caffo:1998du,Laporta:2004rb,Groote:2005ay,Groote:2012pa,Bailey:2008ib,MullerStach:2011ru,Adams:2013nia,Bloch:2013tra,Remiddi:2013joa,Adams:2014vja,Adams:2015gva,Adams:2015ydq}.
\begin{figure}
\begin{center}
\includegraphics[scale=0.8]{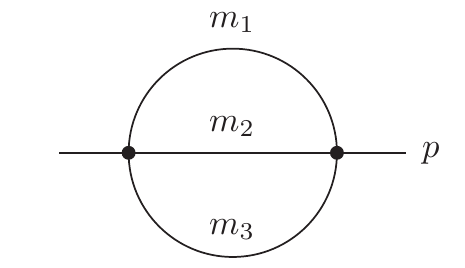}
\caption{\label{fig_sunrise_graph} The two-loop sunset graph.}
\end{center}
\end{figure} 
The two-loop sunset integral, shown in fig.~\ref{fig_sunrise_graph}
is given in $D$-dimensional Minkowski space by
\bq
\label{def_sunrise}
 S_{111}\left( D, p^2, m_1^2, m_2^2, m_3^2, \mu^2 \right)
 & = &
 \left(\mu^2\right)^{3-D}
 \int \frac{d^Dk_1}{i \pi^{\frac{D}{2}}} \frac{d^Dk_2}{i \pi^{\frac{D}{2}}} \frac{d^Dk_3}{i \pi^{\frac{D}{2}}}
 \frac{\delta^D\left(k_1+k_2+k_3-p\right)}{\prod\limits_{i=1}^3\left(-k_i^2+m_i^2\right) }.
\eq
In eq.~(\ref{def_sunrise}) the three internal masses are denoted by $m_1$, $m_2$ and $m_3$. 
The arbitrary scale $\mu$ is introduced to keep the integral dimensionless.
The quantity $p^2$ denotes the momentum squared (with respect to the Minkowski metric) and we will write $t = p^2$.
Where it is not essential we will suppress the dependence on the masses $m_i$ and the scale $\mu$ and simply write
$S_{111}( D, t)$ instead of $S_{111}( D, t, m_1^2, m_2^2, m_3^2, \mu^2)$.
In terms of Feynman parameters the two-loop integral is given by
\bq
\label{def_Feynman_integral}
 S_{111}\left( D, t\right)
 & = & 
 \Gamma\left(3-D\right)
 \left(\mu^2\right)^{3-D}
 \int\limits_{\sigma} \frac{{\cal U}^{3-\frac{3}{2}D}}{{\cal F}^{3-D}} \omega
\eq
with the two Feynman graph polynomials
\bq
 {\cal U} = x_1 x_2 + x_2 x_3 + x_3 x_1,
 & &
 {\cal F} = - x_1 x_2 x_3 t
                + \left( x_1 m_1^2 + x_2 m_2^2 + x_3 m_3^2 \right) {\cal U}.
\eq
The differential two-form $\omega$ is given by
$\omega = x_1 dx_2 \wedge dx_3 + x_2 dx_3 \wedge dx_1 + x_3 dx_1 \wedge dx_2$.
The integration is over
$\sigma = \{ [ x_1 : x_2 : x_3 ] \in {\mathbb P}^2 | x_i \ge 0, i=1,2,3 \}$.

\subsection{The sunset integral viewed from algebraic geometry}

\begin{figure}
\begin{center}
\includegraphics[scale=0.55]{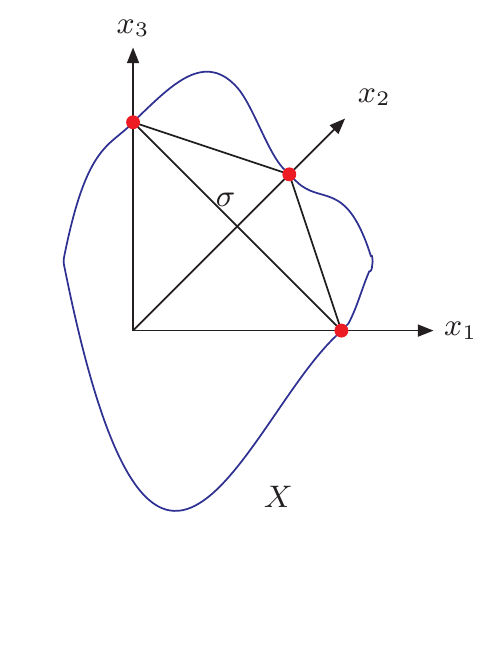}
\hspace*{0mm}
\includegraphics[scale=0.55]{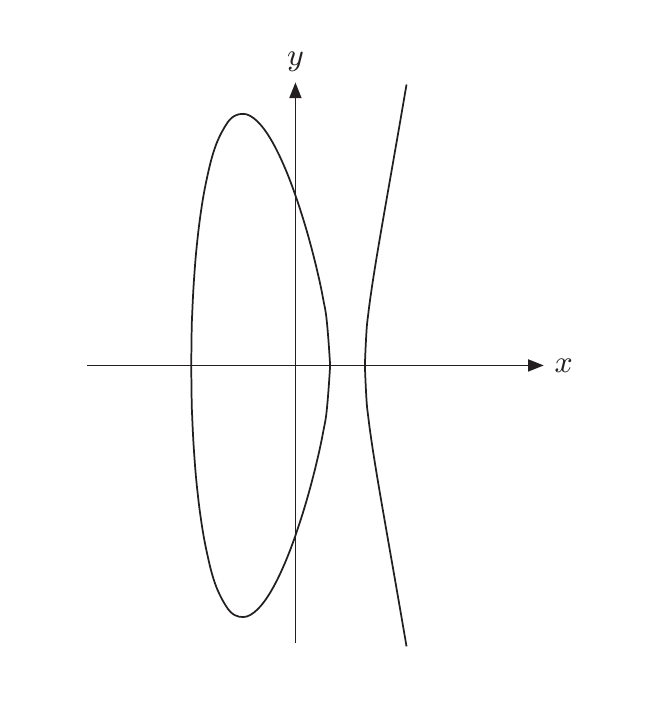}
\hspace*{0mm}
\includegraphics[scale=0.55]{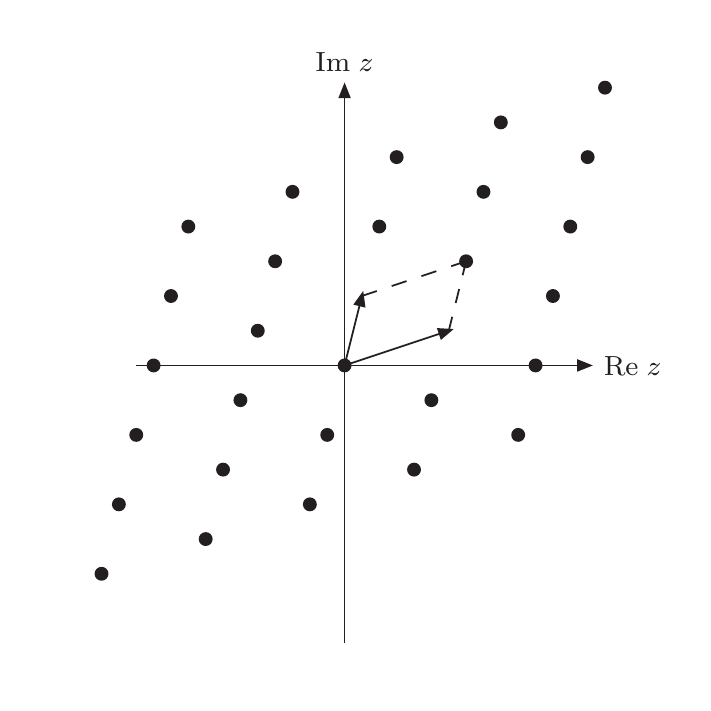}
\hspace*{0mm}
\includegraphics[scale=0.55]{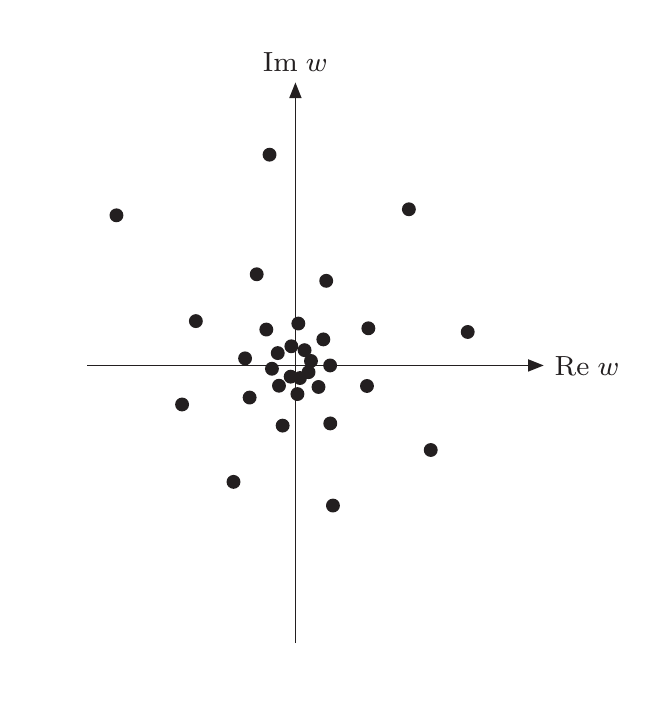}
\caption{\label{fig_elliptic_curve} 
Representations of an elliptic curve:
The left picture shows  a sketch of the integration region $\sigma$ and the zero set $X$ of the graph polynomial ${\mathcal F}=0$. These two sets intersect in three points (red dots) on the coordinate axes.
The next picture shows a plot of an elliptic curve in Weierstrass normal form $y^2=4x^3-g_2x-g_3$.
The third picture shows the torus ${\mathbb C}/\Lambda$, where the lattice
$\Lambda$ is spanned by the two periods $\psi_1$ and $\psi_2$.
Points in ${\mathbb C}$, which differ by a lattice vector, are identified.
The right picture shows the Jacobi uniformization ${\mathbb C}^\ast/q^{2{\mathbb Z}}$, where points of ${\mathbb C}^\ast$
are identified, if they differ by a power of $q^2$.
}
\end{center}
\end{figure}

The sunset integral is finite 
in two space-time dimensions and eq.~(\ref{def_Feynman_integral}) reduces for $D=2$ to
\bq
\label{def_Feynman_integral_2D}
 S_{111}\left( 2, t\right)
 & = & 
 \mu^2
 \int\limits_{\sigma} \frac{\omega}{{\cal F}}.
\eq
The integrand of eq.~(\ref{def_Feynman_integral_2D}) depends only on the graph polynomial ${\mathcal F}$, but
not on the other graph polynomial ${\mathcal U}$.
From the point of algebraic geometry there are only two objects in the game: The region of integration $\sigma$
and the zero set $X$ of the graph polynomial ${\mathcal F}=0$.
These two sets intersect in three points on the coordinate axes. This is shown in the left picture of fig.~\ref{fig_elliptic_curve}.
The equation
\bq
 - x_1 x_2 x_3 t
                + \left( x_1 m_1^2 + x_2 m_2^2 + x_3 m_3^2 \right) \left( x_1 x_2 + x_2 x_3 + x_3 x_1 \right)
 & = & 0
\eq
defines a cubic curve in ${\mathbb P}^2$ (a Riemann surface of genus one)
and -- together with the choice of a rational point as origin --
an elliptic curve.
By a change of coordinates we can bring the elliptic curve into the 
Weierstrass normal form $y^2 z - 4 x^3 + g_2(t) x z^2 + g_3(t) z^3 = 0$.
In the chart $z=1$ this reduces to
$y^2 - 4 x^3 + g_2(t) x + g_3(t) = 0$.
Note that the elliptic curve varies with $t$.
The second picture of fig.~\ref{fig_elliptic_curve} shows a plot of an elliptic curve in Weierstrass normal form.

Away from $D=2$ the other graph polynomial ${\mathcal U}$ will contribute to the integrand.
The set ${\mathcal U}=0$ defines a Riemann surface of genus zero.
We will start the discussion of the analytic result for the sunset integral for $D=2$ and come back to 
the $D \neq 2$-case in section~\ref{sect:recent_results}.

\subsection{The differential equation}

In two dimensions we have a second-order differential equation \cite{MullerStach:2011ru}:
\bq
 \left[ p_2(t) \frac{d^2}{d t^2} + p_1(t) \frac{d}{dt} + p_0(t)  \right] S_{111}\left(2,t\right) 
 & = & \mu^2 p_3(t).
\eq
The order of the differential equation follows from the fact, that the first cohomology group of an elliptic curve is two-dimensional.
The coefficients $p_0$, $p_1$, $p_2$ and $p_3$ are polynomials in $t$.
The explicit expressions can be found in \cite{MullerStach:2011ru}. For illustration purposes let us quote
the explicit results for the equal mass case:
\bq
 \left[ t \left(t-m^2\right) \left(t-9m^2\right)  \frac{d^2}{d t^2} 
        + \left(3t^2-20tm^2 +9m^4\right) \frac{d}{dt} + t -3 m^2  \right] S_{111}\left(2,t\right) 
 & = & -6 \mu^2.
\eq

\subsection{Solutions of the homogeneous differential equation}

As a first step towards the solution of the differential equation we need the solutions of the corresponding 
homogeneous differential equation:
\bq
 \left[ p_2(t) \frac{d^2}{d t^2} + p_1(t) \frac{d}{dt} + p_0(t)  \right] S_{111}\left(2,t\right) 
 & = & 
 0.
\eq
The solutions of the homogeneous differential equation are the periods of the elliptic curve \cite{Adams:2013nia}.
In detail, these solutions are given as follows:
We start from the cubic curve ${\mathcal F}=0$ and pick one of the three points
\bq
\label{intersection_F_sigma}
 P_1 = \left[1:0:0\right], 
 \;\;\;
 P_2 = \left[0:1:0\right], 
 \;\;\;
 P_3 = \left[0:0:1\right]
\eq
as origin $O$ of the elliptic curve. 
By a change of variables we bring this curve into the Weierstrass normal form
$y^2 = 4 (x-e_1)(x-e_2)(x-e_3)$,
with 
$e_1+e_2+e_3=0$.
All three choices will lead to the same Weierstrass normal form.
The explicit expressions for the roots $e_1$, $e_2$ and $e_3$ are
\bq
\label{def_roots}
 e_{1/2}
 =
 \frac{1}{24 \mu^4} \left( -t^2 + 2 M_{100} t + \Delta \pm 3 \sqrt{D} \right),
 & &
 e_3 = - e_1 - e_2,
\eq
with $M_{100}=m_1^2+m_2^2+m_3^2$, $\Delta = \mu_1 \mu_2 \mu_3 \mu_4$, 
$D = ( t - \mu_1^2 )( t - \mu_2^2 )( t - \mu_3^2 )( t - \mu_4^2 )$.
Here, we denote by $\mu_1$, $\mu_2$ and $\mu_3$ the pseudo-thresholds
$\mu_1 = m_1+m_2-m_3$,
$\mu_2 = m_1-m_2+m_3$,
$\mu_3 = -m_1+m_2+m_3$,
and by $\mu_4$ the threshold
$\mu_4 = m_1+m_2+m_3$.
The modulus $k$ and the complementary modulus $k'$ of the elliptic curve are given by
\bq
\label{def_modulus}
 k = \sqrt{\frac{e_3-e_2}{e_1-e_2}},
 & &
 k' = \sqrt{1-k^2} = \sqrt{\frac{e_1-e_3}{e_1-e_2}}.
\eq
The periods of the elliptic curve (and the solutions of the homogeneous differential equation) are then
\bq
\label{def_periods}
 \psi_1 =  
 2 \int\limits_{e_2}^{e_3} \frac{dx}{y}
 =
 \frac{4 \mu^2}{D^{\frac{1}{4}}} K\left(k\right),
 & &
 \psi_2 =  
 2 \int\limits_{e_1}^{e_3} \frac{dx}{y}
 =
 \frac{4 i \mu^2}{D^{\frac{1}{4}}} K\left(k'\right).
\eq

\subsection{The inhomogeneous solution}

Let us now turn to the solution of the inhomogeneous differential equation.
We have to address which transcendental functions we encounter there, and the arguments of these functions.

\subsubsection{Functions in the inhomogeneous solution}
\label{sect:functions}

In addition to the multiple polylogarithms we will need the following transcendental functions
\bq
\lefteqn{
 \mathrm{ELi}_{n_1,...,n_l;m_1,...,m_l;2o_1,...,2o_{l-1}}\left(x_1,...,x_l;y_1,...,y_l;q\right) 
 = }
 & & \nonumber \\
 & = &
 \sum\limits_{j_1=1}^\infty ... \sum\limits_{j_l=1}^\infty
 \sum\limits_{k_1=1}^\infty ... \sum\limits_{k_l=1}^\infty
 \;\;
 \frac{x_1^{j_1}}{j_1^{n_1}} ... \frac{x_l^{j_l}}{j_l^{n_l}}
 \;\;
 \frac{y_1^{k_1}}{k_1^{m_1}} ... \frac{y_l^{k_l}}{k_l^{m_l}}
 \;\;
 \frac{q^{j_1 k_1 + ... + j_l k_l}}{\prod\limits_{i=1}^{l-1} \left(j_i k_i + ... + j_l k_l \right)^{o_i}}.
\eq
We call the linear combination
\bq
 \mathrm{E}_{2;0}\left(x;y;q\right)
 & = &
 \frac{1}{i}
 \left[
 \frac{1}{2} \mathrm{Li}_2\left( x \right) 
 - \frac{1}{2} \mathrm{Li}_2\left( x^{-1} \right)
 + \mathrm{ELi}_{2;0}\left(x;y;q\right)
 - \mathrm{ELi}_{2;0}\left(x^{-1};y^{-1};q\right)
 \right]
\eq
the elliptic dilogarithm \cite{Bloch:2013tra,Adams:2014vja,Bloch:2014qca}.

\subsubsection{Arguments of these functions}

The sunset integral defined in eq.~(\ref{def_sunrise}) 
depends for a given space-time dimension $D$ 
on the variables $t$, $m_1^2$, $m_2^2$, $m_3^2$ and $\mu^2$.
It is clear from the definition that the sunset integral will not change the value under a simultaneous
rescaling of all five quantities.
This implies that the integral depends only on the four dimensionless ratios $t/\mu^2$,
$m_1^2/\mu^2$, $m_2^2/\mu^2$ and $m_3^2/\mu^2$.
It will be convenient to view the non-trivial part of the integral as a function of five new variables
\bq
\label{set_new_variables}
 q, \;\; w_1, \;\; w_2, \;\; w_3 \;\; \mbox{and} \;\; \frac{m_1^2 m_2^2 m_3^2}{\mu^6}.
\eq
The variables $w_1$, $w_2$ and $w_3$ satisfy $w_1 w_2 w_3 = 1$, therefore there are again only four independent variables.
The variables $q$, $w_1$, $w_2$ and $w_3$ are closely related 
to the elliptic curve defined by ${\mathcal F}=0$.

The nome $q$ is defined by
\bq
\label{def_nome}
 q & = & e^{i\pi \tau},
\eq
where $\tau$ is the ratio of the two periods $\psi_2$ and $\psi_1$, given by
$\tau = \frac{\psi_2}{\psi_1}$.
The geometric interpretation of the variables $w_1$, $w_2$ and $w_3$ is as follows:
An elliptic curve can be represented in several ways.
We started from the cubic curve ${\mathcal F}=0$ 
together with the choice of one of the points in eq.~(\ref{intersection_F_sigma}) as origin
and encountered already the Weierstrass normal form.
In addition we may represent an elliptic curve as a torus ${\mathbb C}/\Lambda$, where the lattice
$\Lambda$ is spanned by the two periods $\psi_1$ and $\psi_2$.
Furthermore, there is the Jacobi uniformization ${\mathbb C}^\ast/q^{2{\mathbb Z}}$, where points of ${\mathbb C}^\ast$
are identified, if they differ by a power of $q^2$.
These representations are shown in fig.~\ref{fig_elliptic_curve}.
Recall that we choose one point from eq.~(\ref{intersection_F_sigma}) as origin of the elliptic curve. 
For a given choice there are two points, which are not chosen as origin. We may now look at the images
of these points in the Jacobi uniformization.
Repeating this for all three possible choices as origin, defines six points
$w_1, w_2, w_3, w_1^{-1}, w_2^{-1}, w_3^{-1}$ in the Jacobi uniformization.
In formulas we have
\bq
\label{def_arguments_w_i}
 w_i 
 = 
 e^{i \beta_i},
 \;\;\;\;
 \beta_i 
 = 
 \pi \frac{F\left(u_i,k\right)}{K\left(k\right)},
 \;\;\;\;
 u_i 
 = 
 \sqrt{\frac{e_1-e_2}{x_{j,k}-e_2}},
 \;\;\;\;
 x_{j,k} 
 = 
 e_3 + \frac{m_j^2 m_k^2}{\mu^4}.
\eq
In the definition of $u_i$ we used the convention that $(i,j,k)$ is a permutation of $(1,2,3)$.
In the definition of $\beta_i$ the incomplete elliptic integral of the first kind appears, 
defined by
\bq
 F\left(z,x\right)
 & = &
 \int\limits_0^z \frac{dt}{\sqrt{\left(1-t^2\right)\left(1-x^2t^2\right)}}.
\eq
In the equal mass case we have
$w_1 = w_2 = w_3 = \exp(2\pi i/3)$.

\subsection{Recent results}
\label{sect:recent_results}

Putting everything together, we obtain for the sunset integral 
in two space-time dimensions with arbitrary masses \cite{Adams:2014vja}:
\bq
 S_{111}\left(2,t\right) =
 \underbrace{\vphantom{\frac{4}{\left[\left(\frac{t_1^2}{s_3^4}\right)\right]^{\frac{1}{4}}}} \frac{4}{\left[\left( t - \mu_1^2 \right) \left( t - \mu_2^2 \right) \left( t - \mu_3^2 \right) \left( t - \mu_4^2 \right)\right]^{\frac{1}{4}}}}_{\mbox{algebraic prefactor}} 
 \;\;
 \underbrace{\vphantom{\frac{4}{\left[\left(\frac{t_1^2}{s_3^4}\right)\right]^{\frac{1}{4}}}} \frac{K\left(k\right)}{\pi}}_{\mbox{elliptic integral}}
 \;\;
 \underbrace{\vphantom{\frac{4}{\left[\left(\frac{t_1^2}{s_3^4}\right)\right]^{\frac{1}{4}}}} \sum\limits_{j=1}^3 \mathrm{E}_{2;0}\left(w_j;-1;-q\right)}_{\mbox{elliptic dilogarithms}}.
 \nonumber
\eq
Here, $t$ denotes the momentum squared,
$\mu_1, \mu_2, \mu_3$ the pseudo-thresholds,
$\mu_4$ the threshold,
$K(k)$ the complete elliptic integrals of the first kind,
$k, q$ the modulus and the nome of the elliptic curve,
$\mathrm{E}_{2;0}(x;y;q)$ the elliptic dilogarithm
and $w_1, w_2, w_3$ points in the Jacobi uniformization of the elliptic curve.
The result consists of three parts, an algebraic prefactor, an elliptic integral normalised to $\pi$ and elliptic dilogarithms.

\subsubsection{The sunset integral in $D=4-2\eps$ dimensions}

Up to now we considered the sunset integral in $D=2$ dimensions. Away from $D=2$ dimensions the sunset integral
will depend not only on the graph polynomial ${\mathcal F}$, but also on the graph polynomial ${\mathcal U}$.
Around $D=4-2\eps$ we have the Laurent expansion
\bq
  S_{111}\left( 4-2\eps, t\right)
 & = &
 e^{-2 \gamma \eps} \left[ \frac{1}{\eps^2} S^{(-2)}_{111}(4,t) + \frac{1}{\eps} S^{(-1)}_{111}(4,t) + S^{(0)}_{111}(4,t)
 + {\cal O}\left(\eps\right) \right].
\eq
Around $D=2-2\eps$ we have the Taylor expansion
\bq
  S_{111}\left( 2-2\eps, t\right)
 & = &
 e^{-2 \gamma \eps} \left[ S^{(0)}_{111}(2,t) + \eps S^{(1)}_{111}(2,t)
 + {\cal O}\left(\eps^2\right) \right].
\eq
The pole terms $S^{(-2)}_{111}(4,t)$ and $S^{(-1)}_{111}(4,t)$ are well known and involve only logarithms.
Dimensional recurrence relations relate $S^{(0)}_{111}(4,t)$ to $S^{(0)}_{111}(2,t)$ and $S^{(1)}_{111}(2,t)$.
(In the equal mass case the dependence of $S^{(0)}_{111}(4,t)$ on $S^{(1)}_{111}(2,t)$ drops out.)
The analytic result for $S^{(1)}_{111}(2,t)$ involves only the functions discussed in section~\ref{sect:functions}
and all arguments for the $x$'s and the $y$'s are from the set \cite{Adams:2015gva}
\bq
 \left\{ w_1, w_2, w_3, w_1^{-1}, w_2^{-1}, w_3^{-1}, 1, -1 \right\}.
\eq

\subsubsection{The all-order result in the equal mass case}

In the equal mass case we may consider the full Taylor expansion around $D=2-2\eps$:
\bq
\label{expansion_2D}
  S_{111}\left( 2-2\eps, t\right)
 & = &
 e^{-2 \gamma \eps} \sum\limits_{j=0}^\infty \eps^j S_{111}^{(j)}(2,t).
\eq
Each term in this Taylor expansion can
be expressed in terms of the functions discussed in section~\ref{sect:functions}
and all arguments for the $x$'s and the $y$'s are from the set \cite{Adams:2015ydq}
\bq
 \left\{ e^{\frac{2\pi i}{3}}, e^{-\frac{2\pi i}{3}}, 1, -1 \right\}.
\eq

\section{Summary}

The sunset integral is the simplest Feynman integral, which cannot be expressed in terms of multiple polylogarithms.
It serves as a guide to explore the class of functions beyond the multiple
polylogarithms. Methods from algebraic geometry play a prominent role.
Together with parallel developments on cluster algebras \cite{Golden:2013xva,Golden:2014xqa,Parker:2015cia}
and string amplitudes \cite{Broedel:2014vla,Broedel:2015hia,D'Hoker:2015qmf}
we look at exciting times ahead of us.

\bibliography{/home/stefanw/notes/biblio}
\bibliographystyle{/home/stefanw/latex-style/h-physrev5}

\end{document}